\documentclass[12pt]{article} \usepackage{dina4p} \usepackage{my}
\usepackage{epsfig,amsmath} \usepackage{graphics}
 
\def\lsim{\mathrel{\rlap{\lower4pt\hbox{\hskip1pt$\sim$}}
    \raise1pt\hbox{$<$}}}                
\def\gsim{\mathrel{\rlap{\lower4pt\hbox{\hskip1pt$\sim$}}
    \raise1pt\hbox{$>$}}}                
\newcommand{\cA}{{\cal A}}
\newcommand{\as}{\alpha_\mathrm{s}}
\newcommand{\asb}{{\bar \alpha}_\mathrm{s}}
\newcommand{\ga}{\gamma}
\newcommand{\om}{\omega}

\newcommand{\tqn}{{q}_{t\;n}}
\newcommand{\Pmax}{\bar{q}}
\newcommand{\kt}{k_{t}}
\newcommand{\ktp}{k_{t}^{\prime}}
\newcommand{\SMALLXC}{SMALLXa,SMALLXb}
\newcommand{\CCFM}{CCFMa,CCFMb,CCFMc,CCFMd}
\newcommand{\BFKL}{BFKLa,BFKLb,BFKLc}
\newcommand{\LDC}{LDCa,LDCb,LDCc,LDCd}
\newcommand{\alphasb}{\bar{\alpha}_s}
\newcommand{\JETSET}{Jetsetnew}
\newcommand{\LEPTO}{Ingelman_LEPTO65}
\newcommand{\PYTHIA}{Jetsetc}
\newcommand{\RAPGAPMC}{RAPGAP206}
\newcommand{\DGLAP}{DGLAPa,DGLAPb,DGLAPc,DGLAPd}
\def\CASCADE{{\sc Cascade}}
\def\SMALLX{{\sc Smallx}}
\def\RAPGAP{{\sc Rapgap}}
\begin{document}
\begin{flushright}
 CERN--TH/2000--318\\
 DESY 00-151\\
 LUNFD6/(NFFL--7189) 2000 \\
 LPTHE--00--43\\
 hep-ph/yymmnnn\\
 December 2000\\
\end{flushright}

\vspace*{10mm}
\begin{center}  \begin{Large} \begin{bf}
{\renewcommand{\thefootnote}{\fnsymbol{footnote}}
Hadronic final state predictions from CCFM: \\
the hadron-level Monte Carlo generator \CASCADE\footnote{Research
  supported in part by E.U. QCDNET contract FMRX-CT98-0194.}}\\
  \end{bf}  \end{Large}
  \vspace*{5mm}
  \begin{large}
 H. Jung$^1$, G.P. Salam$^{2,3}$ \vspace{0.3cm}\\
  \end{large}{\small
$^1$ Physics Department, Lund University, Box 118, S-221~00 Lund, Sweden\\  
$^2$ CERN, TH Division, Geneva, Switzerland\\  
$^3$ LPTHE, Universit\'es P. \& M. Curie (Paris VI) et Denis Diderot
(Paris VII), Paris, France \\  }
\end{center}
\setcounter{footnote}{0}
\begin{quotation}
\noindent
{\bf Abstract:} We discuss a practical formulation of backward
evolution for the CCFM small-$x$ evolution equation and show results
from its implementation in the new Monte Carlo event-generator
\CASCADE.
\end{quotation}

\section{Introduction}

In recent years a wealth of experimental results has become available
from HERA concerning structure functions and final-state properties in
deep inelastic collisions (DIS) at small Bjorken $x$ and moderate
$Q^2$ and this has led to interest in theoretical descriptions and
predictions of the phenomena that are observed.

DIS at moderate values of $x$ is well described by resummations of
leading logarithms of transverse momenta $(\as \ln Q^2)^n$, generally
referred to as DGLAP physics \cite{\DGLAP}. At small $x$ we expect
leading-logs of longitudinal momenta, $(\as \ln x)^n$, to become
equally if not more important. However, while the understanding of
DGLAP resummations has been mature for some time now, despite
considerable effort small-$x$ resummations still remain the subject of
many theoretical uncertainties and technical difficulties.

Since many of the measurements at HERA involve complex cuts and
multi-particle final states, the ideal form for any theoretical
description of the data is a Monte Carlo event-generator which
embodies small-$x$ resummations, in analogy with event generators such
as LEPTO~\cite{\LEPTO}, PYTHIA \cite{\PYTHIA}, HERWIG
\cite{Herwig,Herwig54} and RAPGAP~\cite{\RAPGAPMC} which embody DGLAP
resummations.

In order to build such an event generator two ingredients are
required.  Firstly one needs to know the underlying parton branching
equation which, when iterated over many branchings, reproduces the
correct leading logarithms. Secondly one has to find an efficient way
of implementing the branching equation into a Monte Carlo event
generator.

As it happens there are several branching equations which are
advocated by various groups as suitable for describing both inclusive
and exclusive properties of small-$x$ DIS. The main purpose of this
article is to discuss how to formulate one of these equations, the
CCFM equation \cite{\CCFM} in a manner suitable for carrying out a
backward evolution.  This is an almost essential requirement if one
wishes to efficiently generate unweighted Monte Carlo events and
modern DGLAP based Monte Carlo generators are always based on backward
evolution approaches \cite{PYTHIAPSa,PYTHIAPSb,MarchWebbBE}. One of
the main results of this paper is that despite the fact that the CCFM
equation is considerably more complicated than the DGLAP equation, it
is possible to cast the backward evolution in a form which looks quite
similar to the normal DGLAP approach.  We then show predictions
obtained from a new Monte Carlo event generator, \CASCADE, which
implements the backward evolution, and compare them with HERA data.

The paper is structured as follows: in section~\ref{sec:WhyCCFM} we
discuss briefly the reasons for choosing the CCFM equation for the
underlying branching. Then in section~\ref{sec:CCFMEquation} we
discuss the CCFM equation itself, and review some details of its
implementation in the forward-evolution Monte Carlo event-generator
\SMALLX\ \cite{\SMALLXC}, which has been used to generate the unintegrated
gluon distribution required by the backward evolution. In
section~\ref{sec:Backward} we discuss the backward evolution itself,
and then in section~\ref{sec:Results} show some results.

\section{Why CCFM?}
\label{sec:WhyCCFM} 

There are three equations which are commonly used for predictions in
small-$x$ DIS: the BFKL equation \cite{\BFKL}, the CCFM equation
\cite{\CCFM} and the LDC equation \cite{\LDC}. 

The BFKL and CCFM approaches are known to reproduce the correct
small-$x$ leading logarithms for the total cross section.  For
final-state properties, the derivation of the BFKL equation is such
that it is not able to guarantee the correctness of the small-$x$
logarithms. On the other hand the derivation of the CCFM equation,
based on the principle of colour coherence, is such that it guarantees
the correct leading logarithms for all final-state observables. (It so
happens that the CCFM equation also gives a correct description of the
final state in the limit $x \to 1$, another region particularly
sensitive to coherence effects, however this is not relevant for our
purposes).

This was the situation until a couple of years ago. Recently however
it was shown that BFKL gives identical leading logarithms of $x$ to
CCFM for all final-state observables
\cite{ForshawSabioVera98,Webber98,Salam99} (the physical reasons for
this are not fully understood).  Since the BFKL equation is quite
simple (compared to CCFM) one might think that this opens the way to a
BFKL-based Monte Carlo for DIS.  But from the point of view of a
\emph{sensible} description of exclusive quantities, it is not just
the leading logarithms which matter. For example with the current
state of the art, one has two options when implementing the BFKL
equation in a event generator for DIS. One possibility is to use $x$
as the evolution variable --- but it turns out that for some
observables this introduces a weak (but pathological) dependence on
one's infrared cutoff in the subleading logarithms of $x$, $\as^p (\as
\ln x)^n$ ($p>0$) \cite{Salam99}. The second way of implementing BFKL,
which resolves this problem, is to use rapidity as the evolution
variable.  But this turns out to be at the expense of introducing
subleading logarithms $(\as \ln^2 Q^2)^n$ which violate
renormalisation group considerations.\footnote{For certain
  restricted applications in hadron-hadron scattering, where BFKL
  generators have been advocated and developed
  \cite{Schmidt,OrrStirlingA,OrrStirlingB}, these double-logarithms
  may not be too important because of the nature of the observables
  studied.} Specifically DGLAP, CCFM and BFKL with $x$ as the
evolution variable all predict that $F_2$ at small $x$ and large $Q^2$
should behave as 
\begin{equation*}
F_2(x,Q^2) \sim \exp\left(2\sqrt{\asb \ln Q \ln 1/x}\right)
\end{equation*}
(where $\asb = \as C_A/\pi$ and we use fixed $\as$ for the purposes of
illustration).  For BFKL with rapidity as the evolution variable one
has to substitute $\ln 1/x$ with the rapidity difference between the
two ends of the chain $\Delta \eta \simeq \ln Q/x$, and this leads to
$F_2$ behaving as
\begin{equation*}
F_2(x,Q^2) \sim \exp\left(2\sqrt{\asb \ln Q \ln 1/x + \asb \ln^2
    Q}\right), 
\end{equation*}
in contradiction to the DGLAP result.

The CCFM equation does not suffer from these problems and so forms a
good basis for an event generator. This does not mean that the CCFM
equation embodies all of our knowledge about small-$x$ branching, in
particular it is known to be incomplete in regions of phase space
where there is collinear or anti-collinear branching. Furthermore it
was discovered in \cite{Salam} that seemingly small modifications of
the equation can lead to big differences in its predictions.
Specifically in a version of the equation without the $1/(1-z)$ part
of the splitting function, the replacement of $\Theta(k_t-q)$ with
$\Theta(k_t-(1-z)q)$ in the non-Sudakov form factor,
eq.~(\ref{non_sudakov}), was instrumental in enabling a fit to $F_2$.
However around the same time the exact next-to-leading logarithmic
(NLL) corrections to small-$x$ evolution, terms $\as (\as \ln x)^n$,
became available \cite{NLLFL,NLLCC}: they state that the power
$\omega$ governing the growth of quantities like the forward-jet cross
section at small $x$ should have an expansion $\omega = 4\ln2 \asb -
18.4 \asb^2 + \ldots$ (for $n_f = 4$). On the other hand the version
of the CCFM equation with the replacement
$\Theta(k_t-q)\to\Theta(k_t-(1-z)q)$ in the non-Sudakov form factor
leads to $\omega = 4\ln2 \asb - (75\pm4) \asb^2 + \ldots$, which is
quite incompatible with the known expansion \cite{NLLFL,NLLCC} and so
this variant of the CCFM equation can be ruled out. The version of the
CCFM equation that is studied here has %
$\omega = 4\ln2 \asb -(9.2\pm0.5) \asb^2 + \ldots$ %
\cite{BMSSunpublished} --- i.e.\ its NLL corrections are somewhat
smaller than the true NLL corrections, however they are at least of
the right order of magnitude.

There exists also a third evolution equation, the Linked Dipole Chain
(LDC) \cite{\LDC}, whose characteristic is that some of the initial
state radiation has been moved into the final state. This has two
consequences --- on one hand the simplifications that ensue mean that
it is quite easy to correctly implement a symmetry between branching
up and down in transverse scale (or equivalently one obtains the same
predictions whether one evolves from the virtual photon or from the
proton). Such a symmetry leads to the implicit inclusion of important
NLL terms associated with large transverse logs in the anti-collinear
limit (terms $\asb^2/(1-\ga)^3$ where $\ga$ is the Mellin variable
conjugate to squared transverse momentum). These terms are entirely
missing from standard LL BFKL, while in CCFM they are present, but
with half the correct coefficient.\footnote{There have been attempts
  to implement the so-called kinematic constraint
  \cite{CCFMa,LDCa,Martin_Sutton}, which in BFKL does lead to the
  correct NLL behaviour for $\ga\to1$.  But in CCFM the situation is
  more subtle and a straightforward inclusion of the kinematic
  constraint still gives the wrong $\ga\to1$ limit.} The second
consequence of the manner in which initial state radiation has been
moved into the final state, is that the LDC has slightly different
small-$x$ leading logs compared to BFKL ($\om \simeq 3.23\asb +
\ldots$ for LDC as compared to $\om \simeq 2.77\asb + \ldots$ for BFKL
and CCFM --- it is possible though to modify LDC so that it has
leading logs which are much closer to the BFKL ones \cite{LDCgoodLL}).
Therefore LDC has both advantages and disadvantages compared to CCFM.
The question of which matters more will almost certainly depend on the
nature of the observable under study.

So there remains much room for further theoretical progress in the
description of small-$x$ evolution. However given the current
situation, one of the more viable options is the CCFM equation, hence
our interest in implementing a practical CCFM-based event generator.

\section{The CCFM evolution equation}
\label{sec:CCFMEquation}

The implementation of CCFM~\cite{\CCFM} parton evolution in the
forward evolution Monte Carlo program \SMALLX~ is described in detail
in~\cite{\SMALLXC}.  Here we only concentrate on the basic ideas and
discuss the new treatment of the non-Sudakov form factor
\cite{smallx_f2,CASCADE}.

\begin{figure}
\begin{center} 
\input{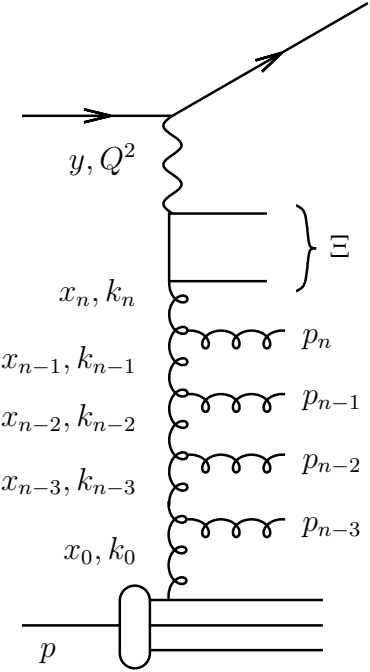}
\end{center} 
\caption{{\it Kinematic variables for multi-gluon emission. The $t$-channel gluon
four-vectors are given by $k_i$ and the gluons emitted in the initial state
cascade have four-vectors $p_i$. The upper angle for any emission is obtained
from the quark box, as indicated with $\Xi$.  
\label{CCFM_variables} }} 
\end{figure}

Figure~\ref{CCFM_variables} shows the pattern of QCD initial-state
radiation in a small-$x$ DIS event, together with labels for the
kinematics.  According to the CCFM evolution equation, the emission of
partons during the initial cascade is only allowed in an
angular-ordered region of phase space. The maximum allowed angle $\Xi$
is defined by the hard scattering quark box,\footnote{Strictly there
  are two maximum angles, corresponding to the directions of the quark
  and the anti-quark, and which each limit roughly one half of the
  radiation.} which connects the gluon to the virtual photon. In terms
of Sudakov variables the quark pair momentum is written as:
\begin{equation}
p_q + p_{\bar{q}} = Y (p_p + \Xi p_e) + Q_t
\end{equation}
where $p_p$ and $p_e$ are the proton and electron momenta,
respectively and $Q_t$ is the transverse momentum of the quark pair.
Similarly, the momenta $p_i$ of the gluons emitted during the initial
state cascade are given by (here treated massless):
\begin{equation}
p_i = y_i (p_p + \xi_i p_e) + p_{ti} \;  , \;\; \xi_i=\frac{p_{ti}^2}{s
y_i^2},
\end{equation}
with $y_i = (1 - z_i) x_{i-1}$ and $x_i = z_i x_{i-1}$ and
$s=(p_p+p_e)^2$ being the total electron proton center of mass energy.
The variable $\xi_i$ is connected to the angle of the emitted gluon
with respect to the incoming proton and $x_i$ and $y_i$ are the
momentum fractions of the exchanged and emitted gluons, while $z_i$ is
the momentum fraction in the branching $(i-1) \to i$ and $p_{ti}$ is
the transverse momentum of the emitted gluon.
\par
The angular-ordered region is then specified by:
\begin{equation}
\xi_0 < \xi_1< \cdots < \xi_n < \Xi
\end{equation}
which becomes:
\begin{equation}
z_{i-1} q_{ti-1} < q_{ti} 
\end{equation}
where we use the rescaled transverse momenta $q_{ti}$ of the emitted
gluons defined by:
\begin{equation}
 q_{ti} = x_{i-1}\sqrt{s \xi_i} = \frac{p_{ti}}{1-z_i}
\end{equation}
\par
In \SMALLX, the initial state gluon cascade is generated 
in a forward evolution approach from an initial distribution of the
$k_t$ unintegrated gluon distribution according to:
\begin{equation}
x {\cal A}_0(x,k_{t0}^2) = N \cdot 5 \frac{1}{k_0^2} 
 (1-x)^4 \cdot \exp{\left(-k_{t0}^2/k_0^2\right)}
\label{eq1}
\end{equation}
where $N$ is a normalization constant.  The exponential distribution
in $k_t^2$ has a width which will be set to $k_0^2 = 1$~GeV$^2$.
The input gluon distribution needs to be adjusted to fit existing
data, but it turns out that the small $x$ behaviour of the structure
function $F_2$ is rather insensitive to the actual choice of $x{\cal
  A}_0(x,k_{t0}^2)$ and only the normalization $N$ acts as a free
parameter, with the constraint:
\begin{equation}
\int x {\cal A}_0(x,k_{t0}^2) dx dk_{t0}^2  = \int x G_0(x,Q^2) dx \simeq 0.5
\end{equation}
which gives:
\begin{equation}
\int x {\cal A}_0(x,k_{t0}^2) dx dk_{t0}^2  = N \cdot 5 \cdot \int
(1-x)^4 dx = N
\end{equation}

\par
The initial state branching of a gluon $k_{i}$ into another virtual
($t$-channel) gluon $k_{i+1}$ and a final gluon $p_{i+1}$ (treated on
mass shell) is generated iteratively from the initial gluon
distribution at a starting scale $Q_0$.  The probability for
successive branchings to occur is given by the CCFM splitting
function~\cite{\CCFM}:
\begin{equation}
d{P}_i  = \tilde{P}_g^i(z_i,q^2_{i},k^2_{ti}) \cdot \Delta_s d z_i
             \frac{d^2 q_{i} }{\pi q^{2}_{i}} 
                 \cdot \Theta(q_{i}-z_iq_{i-1})
                 \cdot \Theta(1-z_i-\epsilon_i)
                 \label{CCFM_splitting}
\end{equation}
with $q_{i}=p_{ti}/(1-z_i)$ being the rescaled transverse momentum of
the emitted gluon $i$. The fractional energy of the exchanged gluon
$i$ is given by $x_i$ and the energy transfer between the exchanged
gluons $i-1$ and $i$ is given by $z_i=x_i/x_{i-1}$. A collinear cutoff
$\epsilon_i=Q_0/q_{i}$ is introduced to regularize the $1/(1-z)$
singularity. The Sudakov form factor $\Delta_s$ is given by:
\begin{equation}
\Delta_s(q_{i},z_iq_{i-1}) =\exp{\left(
 - \int_{(z_{i-1} q_{i-1})^2} ^{q^{2}_{i}}
 \frac{d q^{2}}{q^{2}} 
 \int_0^{1-Q_0/q} dz \frac{\alphasb(q^2(1-z)^2)}{1-z}
  \right)}
  \label{Sudakov}
\end{equation}
with $\alphasb=\frac{C_A \alpha_s}{\pi}=\frac{3 \alpha_s}{\pi}$. For
inclusive quantities at leading-logarithmic order the Sudakov form
factor cancels against the $1/(1-z)$ collinear singularity of the
splitting function.  Coherence effects are taken into account by
angular ordering $q_{i} > z_{i-1} q_{i-1}$ given by the first $\Theta$
function in eq.(\ref{CCFM_splitting}). 
The cascade continues until $q_i$ reaches the limiting angle defined
by $\Pmax= x_{n} \sqrt{s \Xi}$, set by the partons from the hard
scattering matrix element.
\par
The gluon splitting function $\tilde{P}_g^i$ is given
by:\footnote{Actually the `correct' scale for $\as$, as suggested by
  the NLL corrections to BFKL, is probably $q^2_{i}(1-z_i)^2$ in both
  terms, with a corresponding modification of the non-Sudakov form
  factor. However for simplicity, at this stage we have retained the
  scale for $\as$ that was present in the original formulation of
  \SMALLX. }
\begin{equation}
\tilde{P}_g^i= \frac{\alphasb(q^2_{i}(1-z_i)^2)}{1-z_i} + 
\frac{\alphasb(k^2_{ti})}{z_i} \Delta_{ns}(z_i,q^2_{i},k^2_{ti})
\label{Pgg}
\end{equation}
where the non-Sudakov form factor $\Delta_{ns}$ is defined as:
\begin{equation}
\log\Delta_{ns} =  -\alphasb(k^2_{ti})
                  \int_0^1 \frac{dz'}{z'} 
                        \int \frac{d q^2}{q^2} 
              \Theta(k_{ti}-q)\Theta(q-z'q_{ti})
                  \label{non_sudakov}                   
\end{equation}
The principle difference compared to the corresponding DGLAP splitting
function is the appearance of the non-Sudakov form factor
$\Delta_{ns}$, which screens the $1/z$ singularity in eq.(\ref{Pgg}).
It can be expressed as~\cite{Martin_Sutton}:
\begin{equation}
\log\Delta_{ns} = -\asb(k_{ti}^2)
\log\left(\frac{z_0}{z_i}\right)
\log\left(\frac{k^2_{ti}}{z_0z_i q^2_{i}}\right)
\label{ns_new}
\end{equation} 
where
$$z_0 = \left\{ \begin{array}{ll}
              1             & \mbox{if  } k_{ti}/q_{i} > 1 \\
                  k_{ti}/q_{i} & \mbox{if  } z_i < k_{ti}/q_{i} \leq 1 \\
                  z_i             & \mbox{if  } k_{ti}/q_{i} \leq z_i  
                  \end{array} \right. 
$$
which means that in the region $k_{ti}/q_{i} \leq z_i$ we have $\Delta_{ns}=1$,
giving no suppression at all.\footnote{We note that in the original
  version of \SMALLX\ a simplified version of the non-Sudakov form factor was
  used, which however did not give the right answer for $k_t < q$.}

The CCFM equation for the unintegrated gluon density can be
written~\cite{CCFMd,Salam,Martin_Sutton} as an integral equation:
\begin{equation}
{\cal A} (x,\kt,\Pmax ) = {\cal A}_0 (x,\kt,\Pmax ) + \int \frac{dz }{z} 
\int \frac{d^2 q}{\pi q^{2}} \Theta(\Pmax - zq) \Delta_s(\Pmax ,z q) 
\tilde{P}(z,q,\kt) {\cal A}\left(\frac{x}{z},\ktp,q\right) 
\label{CCFM_integral} 
\end{equation}  
with $\ktp = | \vec{k}_{t} + (1-z) \vec{q}|$ and with $\Pmax$ being
the upper scale for the last angle of the emission: $\Pmax > z_n q_n$,
$q_n > z_{n-1} q_{n-1}$, ..., $q_{1} > Q_0$. As before $q$
is used as a shorthand notation for the 2-dimensional vector of the
rescaled transverse momentum $\vec{q}\equiv\vec{q}_t=\vec{p}_t/(1-z)$.  The
splitting function $\tilde{P}(z,q,\kt)$ is defined in eq.(\ref{Pgg})
and the Sudakov form factor $\Delta_s(\Pmax ,z q)$ is given in
eq.(\ref{Sudakov}).

In \cite{CCFMd} a differential form of the CCFM evolution equation is
given, which is obtained from eq.~(\ref{CCFM_integral}) by dividing
both sides by $\Delta_s(\Pmax,Q_0)$ and then differentiating with
respect to $\Pmax$:
\begin{equation}
\Pmax^2\frac{d\; }{d \Pmax^2} 
   \frac{x \cA(x,\kt,\Pmax)}{\Delta_s(\Pmax,Q_0)}=
   \int dz \frac{d\phi}{2\pi}\,
   \frac{\tilde{P} (z,\Pmax/z,\kt)}{\Delta_s(\Pmax,Q_0)}\,
 x'\cA(x',\kt',\Pmax/z) 
\label{CCFM_differential}
\end{equation} 
with $x'=x/z$ and $\kt' = (1-z)/z\vec{q} + \vec{\kt}$ and where
$\vec{q}$ has an azimuthal angle $\phi$. In deriving this equation we
have exploited the fact that the Sudakov form factor can be written as
\begin{equation}
  \Delta_s(\Pmax,zq) = \frac{\Delta_s(\Pmax,Q_0)}{\Delta_s(zq,Q_0)}\,.
\end{equation}
For (\ref{CCFM_differential}) (and the backward evolution formalism
which follows from it) to be correct ${\cal A}_0 (x,\kt,\Pmax )$, must
be of the form
\begin{equation}
  \label{eq:A03var}
  \cA_0 (x,\kt,\Pmax ) = \cA_0 (x,k_t)\, \Delta_s(\Pmax, Q_0)\,.
\end{equation}

\section{Backward evolution: CCFM and \CASCADE}
\label{sec:Backward}

The forward evolution procedure as implemented in \SMALLX\ is a direct
way of solving the CCFM evolution equation including the correct
treatment of the kinematics in each branching. However the forward
evolution is rather time consuming, since in each branching a weight
factor is associated, and only after the initial state cascade has
been generated completely can it be decided whether the kinematics
allow the generation of the hard scattering process. Quite often, a
complete event has to be rejected.

A more efficient procedure to adopt in a full hadron-level Monte Carlo
generator is a backward evolution scheme, analogous to that used in
standard Monte Carlo programs~\cite{PYTHIAPSb,PYTHIAPSa,Herwig} using
a DGLAP type parton cascade.  The idea is to first generate the hard
scattering process with the initial parton momenta distributed
according to the parton distribution functions.  This involves in
general only a fixed number of degrees of freedom, and the hard
scattering process can be generated quite efficiently. Then, the
initial state cascade is generated by going backwards from the hard
scattering process towards the beam particles. In a DGLAP type cascade
the evolution (ordering) is done usually in the virtualities of the
exchanged $t$-channel partons.

According to the CCFM equation the probability of finding a gluon in
the proton depends on three variables, the momentum fraction $x$, the
transverse momentum squared $k_t^2$ of the exchanged gluons and the
maximum angle allowed for any emission 
$\Pmax = x_{n} \sqrt{s \Xi}$.
To calculate this probability, in addition to the details of the
splitting, eq.~(\ref{CCFM_splitting}) 
one needs to know the unintegrated gluon
distribution $\cA(x,k,\Pmax)$ which has to be determined beforehand. 

Given this distribution, the generation of a full hadronic event has
three steps, implemented in a new hadron-level Monte Carlo program,
\CASCADE: 
\begin{itemize}
\item[$\bullet$] 
Firstly, the hard scattering process is generated,
\begin{equation}
\sigma = \int dk_t^2 dx_g {\cal A}(x_g,k_t^2,\Pmax)
 \sigma (\gamma^* g^* \to q \bar{q})\,,
\label{x_section}
\end{equation}
using the off-shell matrix elements given in~\cite[p.~178
ff]{off_shell_me}, with the gluon momentum (in Sudakov
representation):
\begin{equation}
 k = x_g p_p + \bar{x}_g p_e + k_t \simeq x_g p_p  + k_t\,.
\end{equation}
The gluon virtuality is then $-k^2 \simeq k_t^2$.
\item[$\bullet$] The initial state cascade is generated according to
  CCFM in a backward evolution approach (described in the next
  section).
\item[$\bullet$] The hadronisation is performed using the Lund string
  fragmentation implemented in JETSET \cite{\JETSET}.
\end{itemize}

Strictly speaking, before the last step one should also include
angular-ordered final-state radiation from the intial-state gluons.
For simplicity, in this `proof of concept' version of the generator,
this is currently left out.

In the backward evolution there is one difficulty: The gluon
virtuality enters in the hard scattering process and also influences
the kinematics of the produced quarks and therefore the maximum angle
allowed for any further emission in the initial state cascade. This
virtuality is only known after the whole cascade has been generated,
since it depends on the history of the gluon evolution.  In the
evolution equations itself it does not enter, since there only the
longitudinal energy fractions $z_i$ and the transverse momenta are
involved.  This problem can only approximately be overcome by using
$k^2 = k_t^2/(1-x_g)$ for the virtuality which is correct in the case
of no further gluon emission in the initial state.

The Monte Carlo program \CASCADE~ can be used to generate unweighted
full hadron-level events, including initial-state parton evolution
according to the CCFM equation and the off-shell matrix elements for
the hard scattering process. It is suitable both for photo-production
of heavy quarks as well as for deep inelastic scattering.  The typical
time needed to generate one event is $\sim 0.03 $ sec, which is
similar to the time needed by standard Monte Carlo event generators
such as LEPTO~\cite{\LEPTO} or PYTHIA~\cite{\PYTHIA}.  

\subsection{The unintegrated gluon density}
The unintegrated gluon density $x {\cal A}(x,k_{t}^2,\Pmax)$ is
obtained from a forward evolution procedure as implemented in
\SMALLX~\cite{\SMALLXC}. Due to the complicated structure of the CCFM
equation, no attempt is made to parameterize the unintegrated gluon
density. Instead, the gluon density is calculated on a grid in $\log
x$, $\log k_{t}$ and $\log \Pmax$ of $50 \times 50 \times 50$ points
and then linear interpolation is used to obtain the gluon density at
values in between the grid points.
\par
From the initial gluon distribution as used in \SMALLX\ (including the
same collinear cutoff and normalization) a set of values $x_{0\;i}$
and $k_{t0\;i}$ are obtained by evolving up to a given scale $\log
\Pmax$ using the forward evolution procedure of \SMALLX.  This is
repeated $10^7$ times thus obtaining a distribution of the
unintegrated gluon density $x_n {\cal A}(x_n,k_{t\;n}^2,\tqn)$ for the
slice of phase space with a given $\Pmax$ ($\Pmax > \tqn$). To obtain
a distribution in $\log \Pmax$, the above procedure is repeated from
the beginning $50$ times for the different grid points in $\log \Pmax$
up to $\Pmax = 1800$~GeV.

\subsection{Backward evolution formalism}
In the backward evolution we start from the quark box, with an upper
angle given by $\Xi$ and a gluon four-vector $k_n$ (see
Fig.~\ref{CCFM_variables}) and go successively down in the ladder
until we end up at gluon $k_0$.  Thus the first step is to reconstruct
from $\Xi$ and $k_n$ the vectors $q_n=p_n/(1-z_n)$ and $k_{n-1}=k_i$,
with $z_n=x_i/x_n$.  In the next step $k_i,q_n$ play the role of
$k_n,\Xi$ from the first step.  In the further steps $k_{i-1},q_{i}$
play the role of $k_i,q_{i+1}$ and so on, until the gluon $k_0$ is
reached.

The differential form of the evolution equation
eq.(\ref{CCFM_differential}) gives the (non-normalized)
probability~\cite{PYTHIAPSb}, that during a small decrease of $\Pmax$,
a $t$-channel gluon $k'$ with momentum fraction $x'$ becomes resolved
into a $t$-channel gluon $k$ with momentum fraction $x=z x'$ and an
emitted gluon $q_i$.  During a small decrease of $\Pmax$, a gluon $k$
may be unresolved into a gluon $k'$.  The normalized probability for
this to happen is given by
\begin{equation}
  \frac{\Delta_s(\Pmax,Q_0)}{x \cA(x,\kt,\Pmax)}
  \,d\!\left(\frac{x \cA(x,\kt,\Pmax)}{\Delta_s(\Pmax,Q_0)}\right)
 =
\frac{d \Pmax^2}{\Pmax^2}
\int dz \frac{d\phi}{2\pi}\, \tilde{P} (z,\Pmax/z,\kt) \frac{x'{\cal
    A}(x',\kt',\Pmax/z)} 
{x{\cal A} (x,\kt,\Pmax)}
\label{CCFM_differential_bw}
\end{equation}
This equation can be integrated between $\Pmax$ and $q$ giving:
\begin{equation}
\log \left( \frac{ {\cal A} (x,\kt,q) }   
 { {\cal A} (x,\kt,\Pmax)}
 \frac{\Delta_s(\Pmax,Q_0)}{\Delta_s(q,Q_0)}\right) = - 
\int_q ^{\Pmax}\frac{d {q'}^2}{{q'}^2}
\int dz \frac{d\phi}{2\pi}\, \tilde{P} (z,q'/z,\kt) \frac{x'{\cal
    A}(x',\kt',q'/z)} 
{x{\cal A} (x,\kt,q')}
\end{equation}
Thus the probability for no radiation in the angular ordered region
between $\Pmax$ and $q$ is just given by a new effective form factor
for the backward evolution:
\begin{equation}
{\cal P}_{no\;rad}  (\Pmax, q) = \exp{\left(-
\int_q^{\Pmax}\frac{d {q'}^2}{{q'}^2}
\int dz \frac{d\phi}{2\pi}\,\tilde{P} (z,q'/z,\kt) \frac{x'{\cal
    A}(x',\kt',q'/z)} 
{x{\cal A} (x,\kt,q')}
\right)} 
\label{Sudakov_bw}
\end{equation} 
In principle one could equally well just use 
\begin{equation}
  \label{Sudakov_bw_MW}
  {\cal P}_{no\;rad} (\Pmax, q) = \frac{ {\cal A} (x,\kt,q) }   
 { {\cal A} (x,\kt,\Pmax)}
 \frac{\Delta_s(\Pmax,Q_0)}{\Delta_s(q,Q_0)}\,,
\end{equation}
which is more akin to the backward evolution approach of
\cite{MarchWebbBE}. One can see that this is the correct
probability since from eq.~(\ref{CCFM_integral}) it corresponds to the
fraction of $\cA (x,\kt,\Pmax)$ which comes from angles below $q$.
Though eq.~(\ref{Sudakov_bw}) looks more complicated than
eq.~(\ref{Sudakov_bw_MW}), it turns out that the former is numerically
more suited to our particular situation, because it is less sensitive
to imprecisions and irregularities of $\cA(x,k,q)$ (which we recall is
generated by a forward evolution Monte Carlo approach).  The standard
DGLAP backward evolution equation would be obtained from
eq.~(\ref{CCFM_differential_bw}) by setting $\Delta_{ns}=1$ in the
splitting function $\tilde{P}$ and replacing the argument $\Pmax/z$ in
the parton density function in the r.h.s. of
eq.~(\ref{CCFM_differential_bw}) with $\Pmax$.

In the CCFM backward evolution, starting with the gluon $k_n$, we need
to reconstruct the momentum of the next emitted gluon $q_n$ as well as 
that of the next exchanged gluon, $k_{n-1}$. This is done as
follows. We start with $\Pmax = \Xi$. Then
\begin{itemize}
\item[$a$.] We check whether there is any evolution to be done.  The
  probability that the evolution should stop straight away is given by
  \begin{equation}
    \frac{{\cal A}_0(x_{n},k_{t n}^2, \Pmax)}
    {{\cal A}(x_{n},k_{t n}^2, \Pmax)} \,,
  \end{equation}
  i.e.\ the fraction of the unintegrated gluon distribution that comes
  from the initial distribution.
\item[$b$.] If the evolution continues then the quantity $q' \equiv |z_n
  q_{n}|$ is determined by choosing a random number $R$ uniformally
  distributed between $0$ and $1$ and solving ${\cal
    P}_{no\;rad}(\Pmax,q') =R$ for $q'$.
\item[$c$.] The values of $z_n$ and $\phi_n$ are then chosen randomly
  according to the distribution given in the inner integral of
  eq.~\eqref{Sudakov_bw}. In doing so we implicitly have to
  reconstruct $k_{n-1}$ as well.
\end{itemize}
This completes the reconstruction of a single branching. The procedure
is then repeated with $\Pmax = q_n$ and $n \to n-1$, and so on, until
the evolution stops (step $a$).

In practice the numerical calculation of the integrals in the
effective form factor eq.(\ref{Sudakov_bw}) would be too time
consuming, since they involve the non-Sudakov form factor in the
splitting function and also the ratio of the structure functions. A
simple solution to the problem is the veto algorithm described
in~\cite{PYTHIAPSb}.  The essential point is to find a simple
analytically integrable function, which is always larger than the
integrand in eq.(\ref{Sudakov_bw}). The splitting function 
$\tilde{P} (z,q',\kt)$ can be replaced with:
\begin{equation}
P_{gg}^{appr}(z_{i},k_{t i}^2) = \frac{\alphasb(k_{ti})}{z_{i}} +
\frac{\alphasb(q_{t\,min})}{1-z_{i}} 
 \geq \tilde{P}_{gg}
 \label{splitting_simple}
\end{equation}
The limits on $z_{i}$ are given by:
\begin{equation*}
  x_i \leq z_{i} \leq 1 - \frac{x_i Q_0}{q_{i+1}},
\end{equation*}
which is a larger range than the true one: $z_{i} < 1 - Q_0/q_{i}$
(but $q_{i}$ is not determined at this stage).  Next, the structure
functions which appear in the Sudakov form factor are replaced with
their maximum and minimum values for $x$ and $x'>x$. Thus a simple
analytically calculable form of the Sudakov form factor is obtained:
\begin{equation}
{\cal P}^{simple}_{no\;rad} = \exp{\left(-
\int_q ^{\Pmax}\frac{d {q'}^2}{{q'}^2}
\int dz P_{gg}^{appr}(z,k_{t}) \frac{x'{\cal A}_{max}(x>x')}
{x{\cal A}_{min} (x,\kt)} 
\right)} < {\cal P}_{no\;rad}
\label{Sudakov_bw_simple}
\end{equation}
After $q_i$ and $z_i$ is generated, the true limits on $z$ can be
applied: $x_i \leq z_{i} < 1 - Q_0/q_{i}$. If a $z_i$ lies outside the
true region, a new set of $q_i$ and $z_i$ is generated.  Having
generated the branching variables according to $z_i$, $k_{ti-1}$ and
$q_i$ according to eq.(\ref{Sudakov_bw_simple}) and
eq.(\ref{splitting_simple}), a branching is accepted with a
probability according to the ratio of the integrands of ${\cal P}$
(via eq.(\ref{Sudakov_bw})) and ${\cal P}^{simple}_{no\;rad}$ (via
eq.(\ref{Sudakov_bw_simple})), as formulated in the veto
algorithm~\cite{PYTHIAPSb}.

As mentioned already above, the true virtuality of the $t$-channel
gluons can only be reconstructed after the full cascade has been
generated. By going from the last gluon (closest to the proton), which
has virtuality $k_0^2=k_{t0}/(1-x_0)$, forward in the cascade to the
hard scattering process, the true virtualities of the $k_i^2$ are
reconstructed.  At the end, the gluon entering to the quark box will
have a larger virtuality than without initial state cascade. Thus a
check is performed as to whether the production of the quarks is still
kinematically allowed. If not, the whole cascade is rejected, and the
event without the cascade is kept. This typically happens about $1\%$
of the time.

\subsection{Comparison with the forward evolution cascade in \SMALLX }

In this section, the reconstruction of the parton level cascade
obtained in the backward evolution approach, as described above, is
compared to that of the forward evolution in \SMALLX.  At parton
level, both approaches are expected to be identical, but small
differences can occur due to the finite grid size used to define of
the unintegrated gluon density.  As already mentioned the virtuality
of the gluon entering the hard scattering process is only known after
the complete reconstruction of the initial state cascade, which could
also result in small differences to the forward evolution approach.
In the following comparison, we have used $b\bar{b}$ photoproduction at
$\sqrt{s}=300$~GeV.
\begin{figure}[htb]
  \vspace*{2mm}
\epsfig{figure=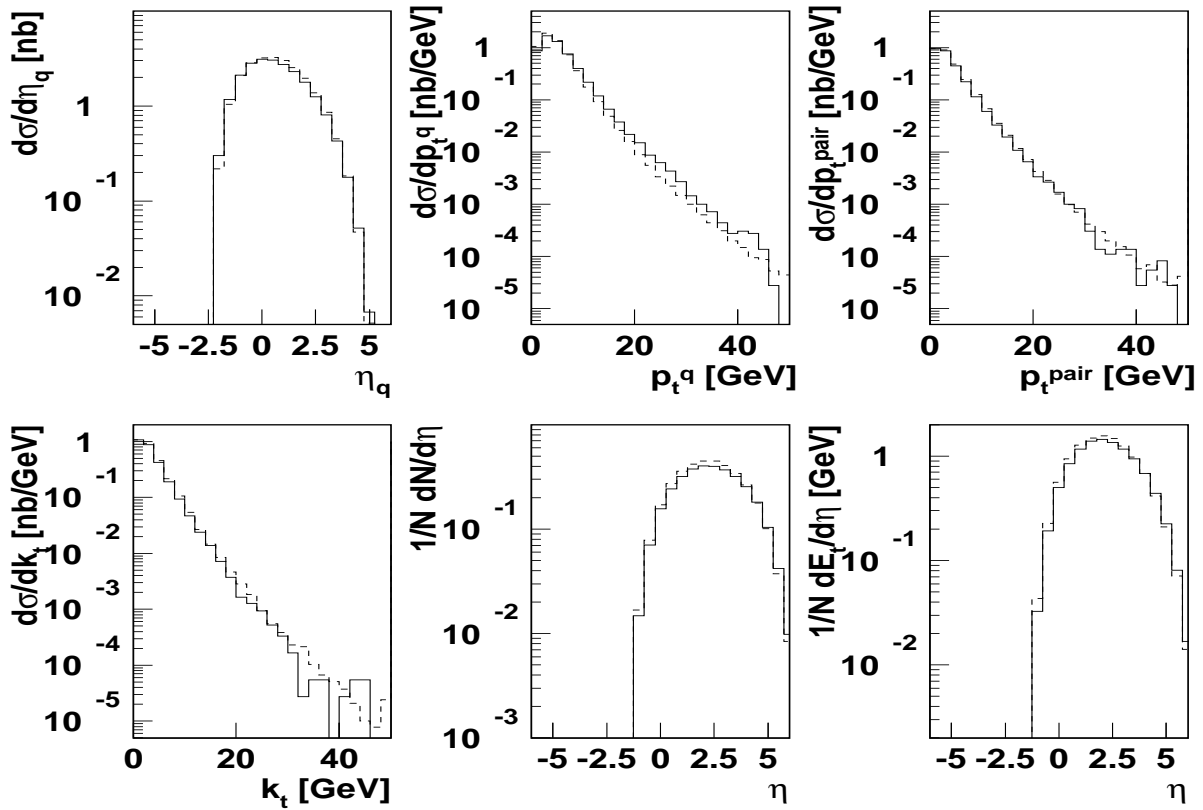,
width=17cm,height=12cm}
\caption{{\it
    Comparison of the cross section obtained from the backward
    evolution Monte Carlo \CASCADE~ (solid line) with \SMALLX~ (dashed
    line) both at parton level only.  The upper plots show the cross
    section as a function of the quark rapidity $\eta_q$, the quark
    transverse momentum $p_t$ and the transverse momentum of the quark
    pair $p_t^{pair}$.  The lower plots show the cross section as a
    function of the gluon transverse momentum $k_t$, and the
    multiplicity and transverse energy flow of the gluons from the
    initial state cascade as a function of the rapidity $\eta$.
    }}\label{smallx_cascade_xsec}
\end{figure}
In Fig.~\ref{smallx_cascade_xsec} the cross section as a function of
the rapidity 
(all rapidities are given in the laboratory frame) 
of the quarks $\eta_q$, the quark transverse momentum $p_t^q$,
the transverse momentum of the quark pair $p_t^{pair}$ and the gluon
transverse momentum $k_t$ obtained from \CASCADE~ (solid line) are
compared to the ones obtained from \SMALLX~ (dashed line).  The
quantities related to the hard scattering matrix element agree very
well.  Also shown in Fig.~\ref{smallx_cascade_xsec} is a comparison of
the multiplicity and the transverse energy flow as a function of
rapidity, and perfect agreement is again found between the backward
and forward evolution approaches.
\begin{figure}[htb]
  \vspace*{2mm}
\epsfig{figure=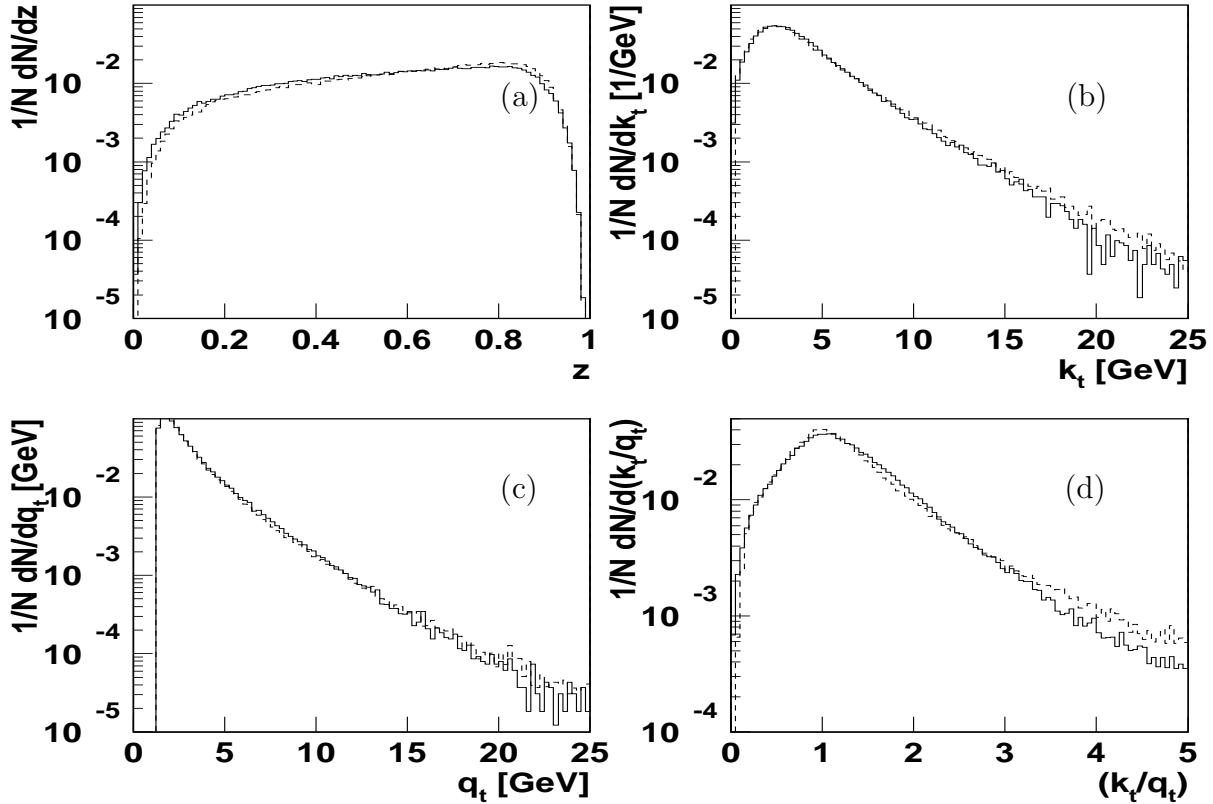,
width=17cm,height=12cm}
\vskip -10.5cm 
\hspace*{7cm}(a)\hspace*{7cm}(b)
\vskip 4.7cm 
\hspace*{7cm}(c)\hspace*{7cm}(d)
\vskip 4.5cm 
\caption{{\it
    Comparison of quantities of the initial cascade obtained from the
    backward evolution Monte Carlo \CASCADE\ (solid line) with
    \SMALLX\ (dashed line) both at parton level only;  $(a)$ shows the
    splitting variable $z$, $(b)$ gives the transverse momentum $k_t$,
    $(c)$ shows the transverse momentum of the emitted gluon $q_t$, and
    $(d)$ shows the ratio $k_t/q_t$.  }}\label{smallx_cascade}
\end{figure}
In Fig.~\ref{smallx_cascade} a more detailed comparison of the
kinematics in the initial state cascade is performed. We compare the
values of the splitting variable $z$, the transverse momenta $k_t$ and
$q_t$ as well as $k_t/q_t$ within the two approaches.

Thus we have described a backward evolution approach, which is fast
and already implemented in the hadron-level Monte Carlo program
\CASCADE, and which reproduces perfectly the parton level
configurations obtained from \SMALLX. This is the first time that a
practical CCFM-based small-$x$ Monte Carlo event generator has been
constructed.

\section{Results}
\label{sec:Results}

In this section we compare predictions from \CASCADE~ with recent
measurements made at HERA. The free parameters of the initial gluon
distribution were fitted to describe the structure function
$F_2(x,Q^2)$ in the range $x<10^{-2}$ and $Q^2 >5$ GeV$^2$.
\begin{figure}[htb]
  \vspace*{2mm}
\epsfig{figure=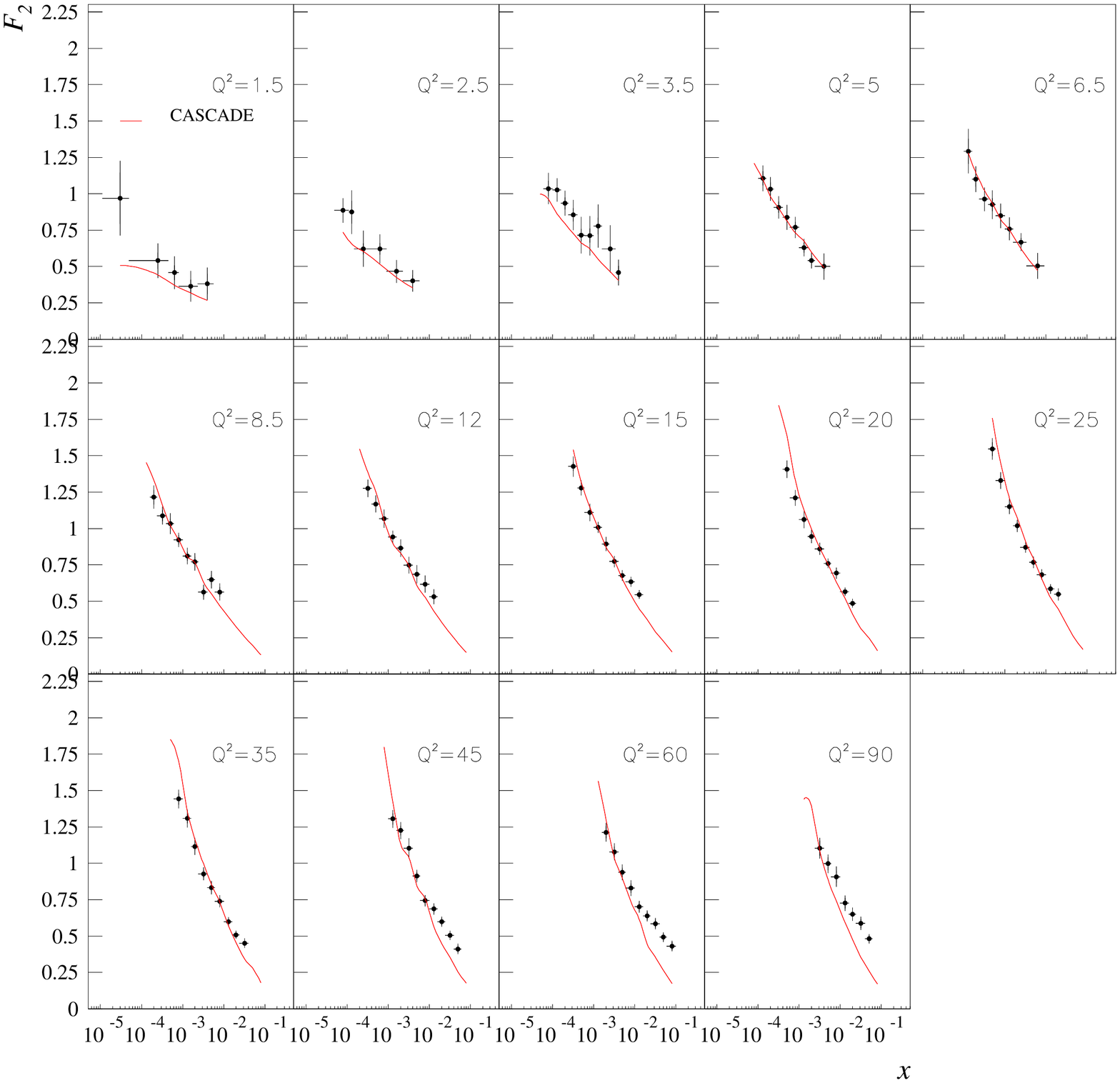,
width=17cm,height=14cm}
\caption{{\it
    Comparison of the structure function $F_2$ as obtained from the
    backward evolution Monte Carlo \CASCADE\ with H1
    data~\protect\cite{H1_F2_1996}.  }}\label{f2cascade}
\end{figure}
The structure function $F_2(x,Q^2)$ as calculated from \CASCADE\ is
compared to a larger range of data in Fig.~\ref{f2cascade}.
Reasonable agreement with the data is observed at small $x$ and $Q^2$.
Deviations from the data are seen at larger $x$ and $Q^2$, which can
be attributed to the quark contributions, which are still missing
here.
\begin{figure}[htb]
  \vspace*{2mm}
  \epsfig{figure=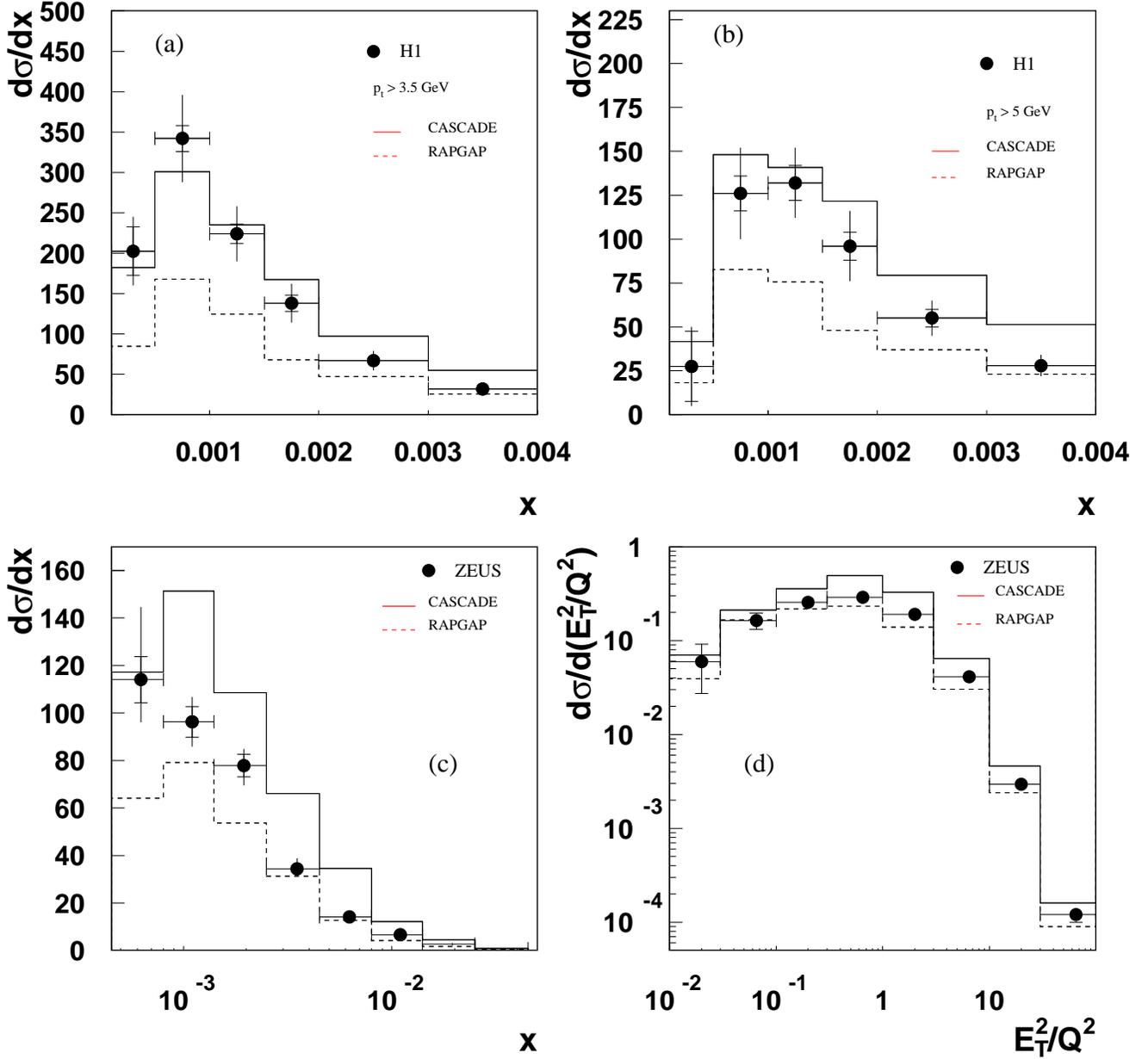,
    width=\textwidth,height=17cm}
\caption{{\it
    The cross section for forward-jet production obtained from the
    Monte Carlo \CASCADE~at hadron level (solid line);  $(a-c)$ The
    cross section for forward-jet production as a function of $x$, for
    different cuts in $p_t$ compared to H1
    data~\protect\cite{H1_fjets_data} ($a-b$) and compared to ZEUS
    data~\protect\cite{ZEUS_fjets_data} ($c$);  $(d)$ The cross
    section for forward-jet production as a function of $E^2_T/Q^2$
    compared to \protect\cite{ZEUS_fjets_pt2/q2}.
    }}\label{fwdjet_cascade}
\end{figure}
\par
In Fig.~\ref{fwdjet_cascade} the cross section predicted for
forward-jet production is shown and compared to measurements done at
HERA. We observe a reasonable description of the data.  However, the
prediction lies above the data, which could indicate that some
relevant sub-leading effects are still missing.
\par
Studies based on a variety of QCD-based Monte Carlos have demonstrated
that the high $p_T$ tail of charged particle transverse momentum
spectra is sensitive to small $x$ dynamics of the parton radiation and
that there is relatively little dependence on the particular model of
hadronisation that is used \cite{Kuhlena,Kuhlenb}.
\begin{figure}[htb]
  \vspace*{2mm}
\epsfig{figure=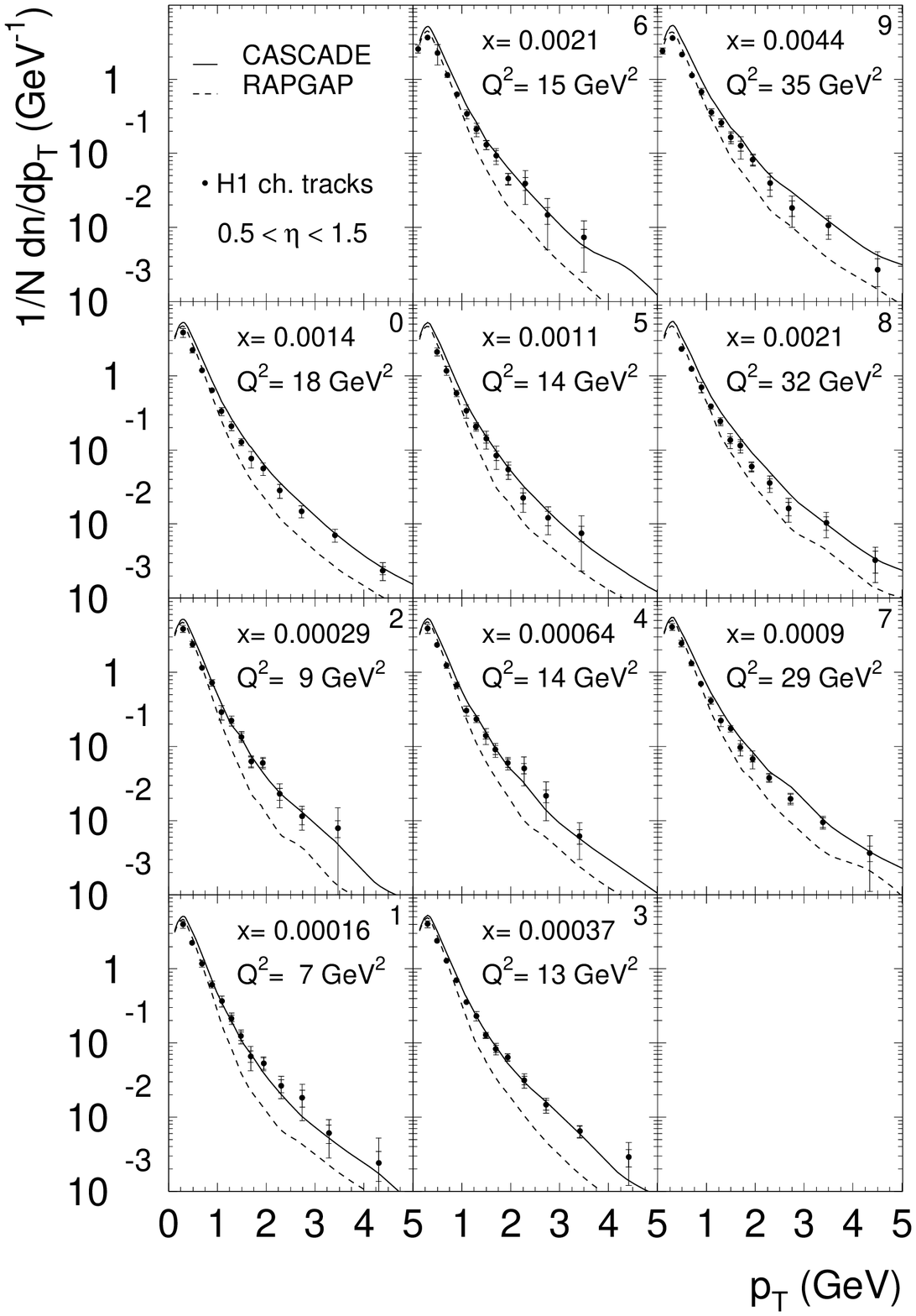,
width=17cm,height=19cm}
\caption{{\it
    The transverse momentum distribution of particles in different
    bins of rapidity.  The prediction from \CASCADE\ (solid line) at
    hadron level is compared to the measurement of
    H1~\protect\cite{H1_ptspectra_data}.  For comparison the
    prediction from the DGLAP based Monte Carlo \RAPGAP\ (dashed line)
    is also shown.  }}\label{particle_spectra}
\end{figure}
Fig.~\ref{particle_spectra} shows the $p_T$ distributions of charged
particles as measured by H1~\cite{H1_ptspectra_data} for DIS
events with 3 GeV$^2 < Q^2 < 70$ GeV$^2$ in the rapidity range $0.5 <
\eta < 1.5$. The prediction from \CASCADE~ gives a good description of
the data, whereas DGLAP based Monte Carlo calculations fail to
describe the data at small $x$ and large $p_t$.

Effects of small $x$ parton dynamics could also show up in
photo-production of charm mesons. Recent
measurements~\cite{ZEUS_dstar} show significant deviations from LO and
NLO QCD calculations and also from hadron-level Monte Carlo
predictions.
\begin{figure}[htb]
\begin{center}
\hbox{
\hspace*{-4.3cm}\vbox{\epsfig{figure=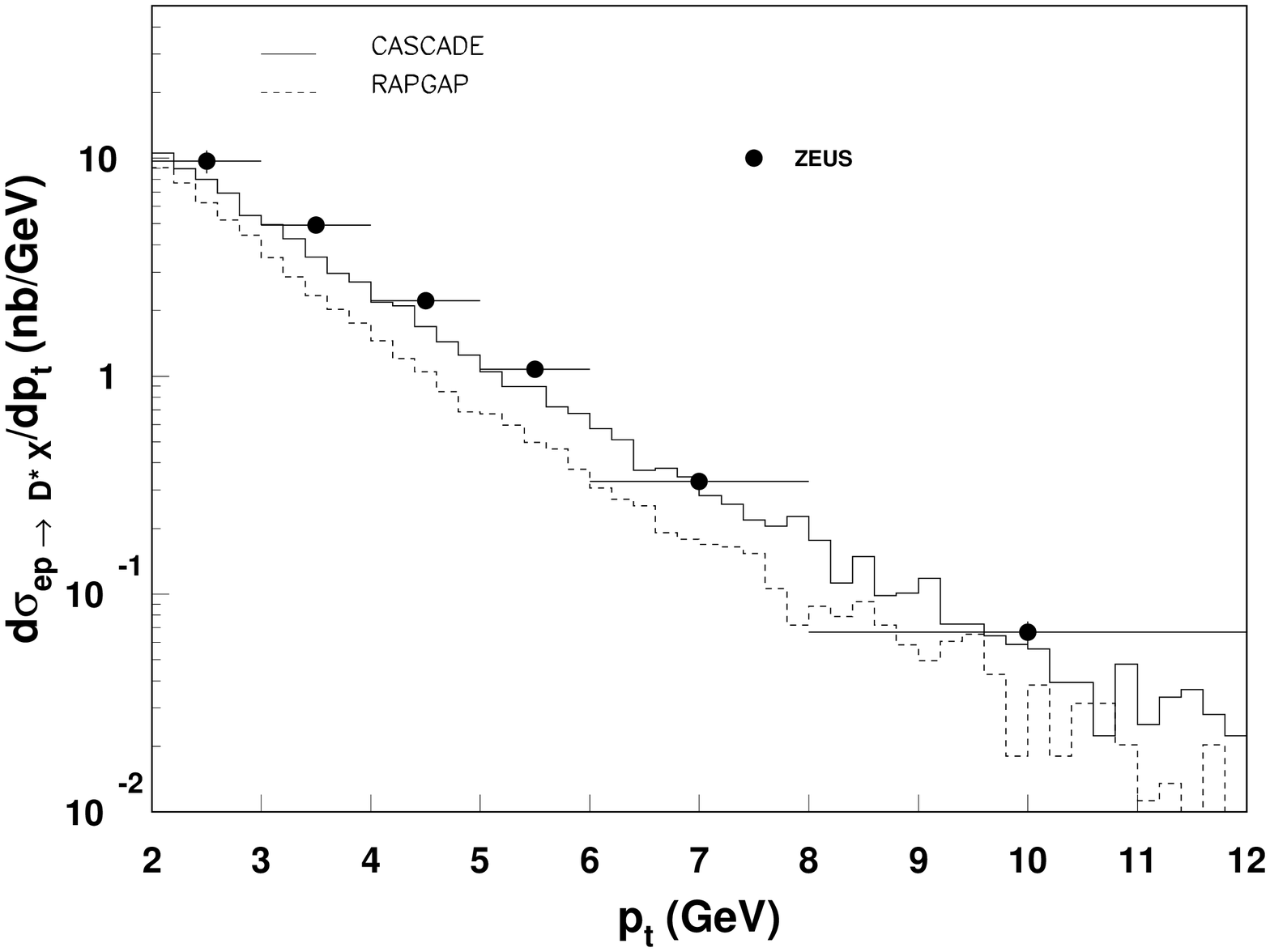,width=9cm,height=9cm}}
\hspace*{-9.0cm}\vbox{\epsfig{figure=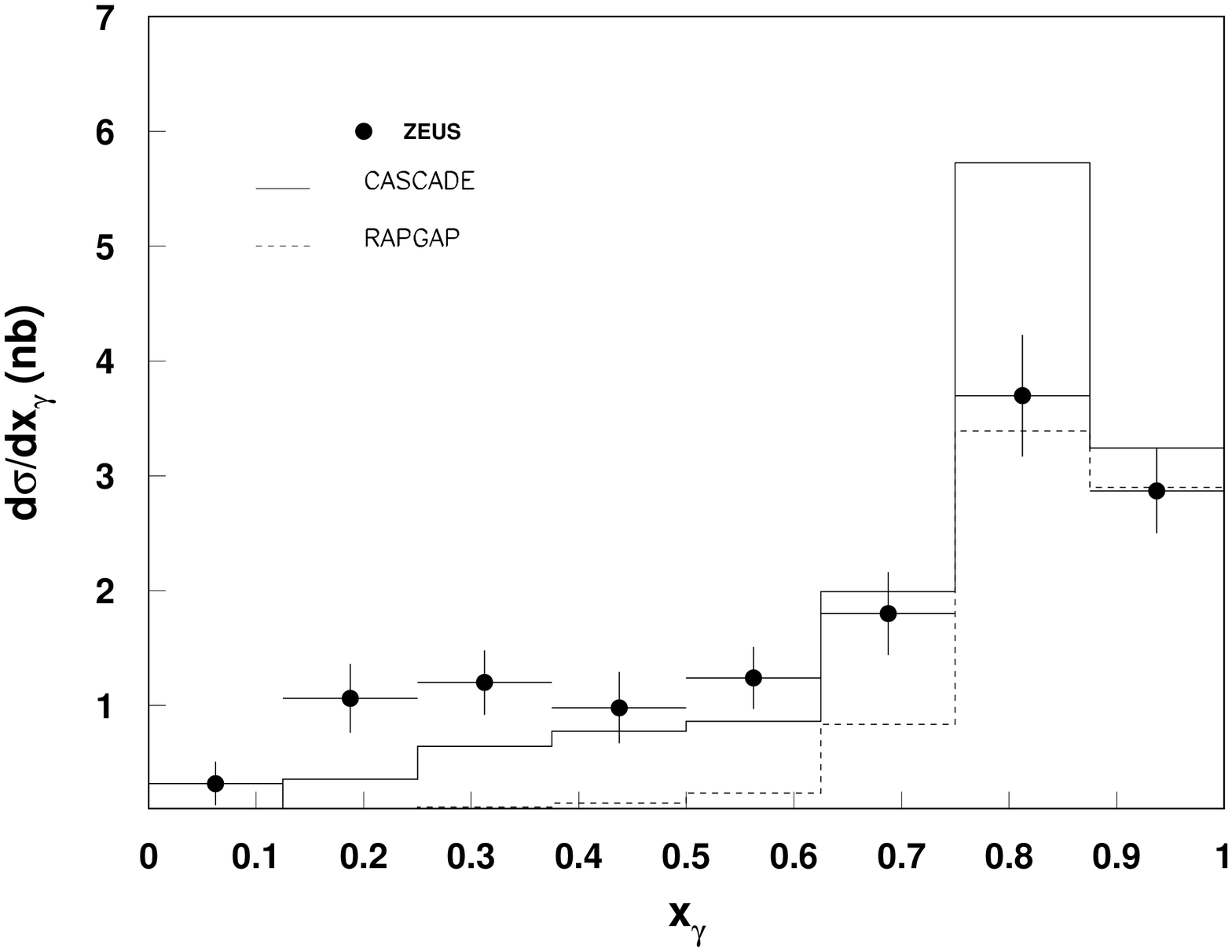,width=9cm,height=9cm}}
}
\end{center}
\vskip -8.5cm 
\hspace*{7cm}(a)\hspace*{7.5cm}(b)
\vskip 7.0cm 
\caption{{\it 
    The differential cross sections of $D^*$ photo-production
    \protect\cite{ZEUS_dstar} for $d \sigma /dp_t^{D^*}$ $(a)$ and $d
    \sigma /dx_{\gamma}$ $(b)$.  The solid line shows the prediction
    from \CASCADE\ while the dashed line shows the prediction from
    \RAPGAP.  }}\label{dstar_ptxgamma}
\end{figure}
In Fig.~\ref{dstar_ptxgamma}$a$ we show the cross section of $D^*$
production as a function of the transverse momentum $p_t^{D^*}$ using
\CASCADE~ and compare it with the measurement of
ZEUS~\cite{ZEUS_dstar}.  For comparison we also show the prediction
from \RAPGAP.  In Fig.~\ref{dstar_ptxgamma}$b$ the $x_{\gamma}$ cross
section is shown.
\begin{figure}[htb]
  \vspace*{2mm}
\epsfig{figure=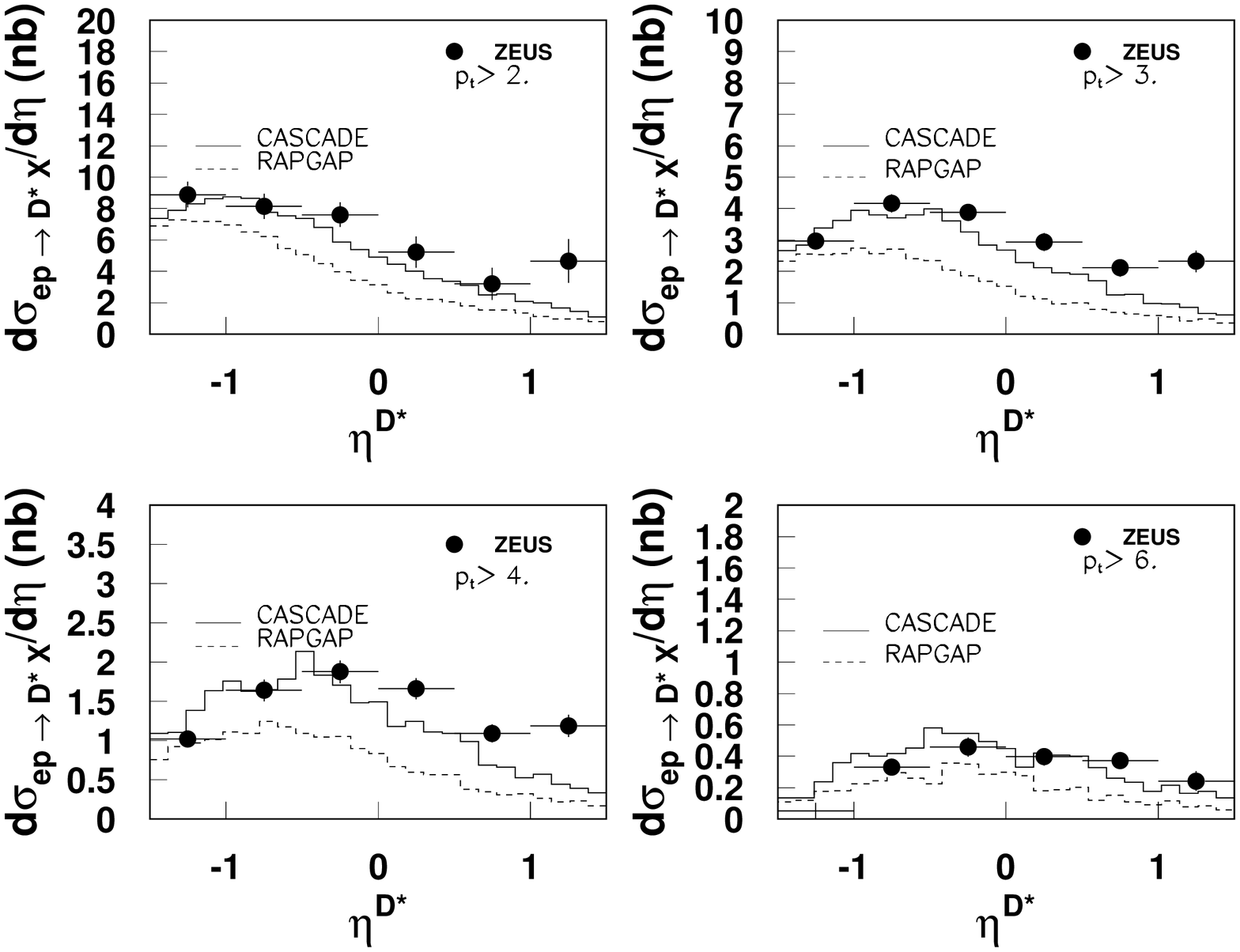,
width=16cm,height=12cm}
\caption{{\it The differential cross section 
    of $D^*$ photo-production \protect\cite{ZEUS_dstar} $d \sigma
    /d\eta^{D^*}$ for different regions of $p_t^{D^*}$.  The solid
    line shows the prediction from \CASCADE\ and the dashed line is
    the prediction from \RAPGAP.  }}\label{dstar_eta}
\end{figure}
In Fig.~\ref{dstar_eta} we show the $D^*$ cross section as a function
of the pseudo-rapidity $\eta^{D^*}$ for different regions in $p_t$.
In all distributions of $D^*$ photo-production we observe a good
description of the experimental data, whereas there is disagreement
between the data and the \RAPGAP\ Monte Carlo predictions.
\par
Also interesting is the cross section for $b\bar{b}$ production as
measured at HERA by H1~\cite{H1_bbar}: $$\sigma(ep \to e' b \bar{b}X)
= 7.1\pm0.6(stat.)^{+1.5}_{-1.3}(syst.)\; \mathrm{nb}$$
to be compared with the
prediction from \CASCADE :
\begin{equation*}
\sigma(ep \to e' b \bar{b}X) = 5.2 ^{+1.1}_{-0.9}\;\mathrm{nb}.
\end{equation*}
The central value is given for $m_b=4.75$~GeV, and the errors are
those associated with a variation of $m_b$ by $\mp 0.25$~GeV.
The NLO calculations predict a cross section which is about a factor
of 2 below the measurements. 
The ZEUS collaboration published a
measurement of $b\bar{b}$ production~\cite{ZEUS_bbar} for the region 
 $p_t>5$~GeV and $|\eta_b| <2$: 
$$\sigma(ep \to e' b \bar{b}X)
= 1.6\pm0.4(stat.)^{+0.3}_{-0.5}(syst.)^{+0.2}_{-0.4}(ext.) 
= 1.6^{+0.54}_{-0.75}\; \mathrm{nb}$$
where we have added all errors quadratically in the last expression. This can
be compared with the
prediction from \CASCADE :
\begin{equation*}
\sigma(ep \to e' b \bar{b}X) = 0.88 \pm 0.08 \;\mathrm{nb}.
\end{equation*}
Again we have used $m_b=4.75 \mp 0.25$~GeV.  The predicted cross
section is still within the total error of the measurement.  The NLO
prediction as given in~\cite{ZEUS_bbar} is
$\sigma=0.64^{+0.15}_{-0.1}$~nb, where the errors are those associated
with a variation of $m_b=4.75 $~GeV by $\mp 0.25 $~GeV and of the
factorisation and renormalisation scales by a factor of $1/2$ and $2$.

\section{Conclusion}
We have shown that a backward evolution approach using the CCFM
evolution equation is possible, and that it works, producing the same
results as the forward evolution used to solve the CCFM equation. The
advantage of backward evolution is the easy implementation into a full
hadron-level Monte Carlo program \CASCADE, which is compared to recent
measurements of small $x$ hadronic final state properties at HERA. We
have found that all small $x$ signatures can be reasonably well
described within one consistent approach. By performing only a fit to
the structure function $F_2(x,Q^2)$ we obtain simultaneously a good
description of a variety of processes which could not be described
within DGLAP: the forward-jet cross section, the high $p_t$ particle
spectra, charm production and a reasonable agreement with the measured
$b\bar{b}$ cross section.  This shows that the CCFM evolution equation
is indeed a good starting point for a consistent description of small
$x$ phenomena.
\section{Acknowledgments}
We are very grateful to B.~Webber for providing us with the \SMALLX~
code, which was the basis of the studies presented here.  We are also
grateful to B.~Andersson, G.~Gustafson, L.~J\"onsson, H.~Kharraziha
and L.~L\"onnblad for all their criticism and useful ideas on CCFM and
the backward evolution, to M.~Ciafaloni and S.~Catani for comments on the
manuscript.  One of us (H.J.) would like to thank A.D.~Martin and
J.~Kwiecinski for their explanation of the modified non-Sudakov form
factor and the DESY directorate for hospitality and support.


\begin{thebibliography}{10}

\bibitem{DGLAPa}
V.~Gribov, L.~Lipatov, {\em Sov. J. Nucl. Phys.} {\bf 15}\nolinebreak
  [2]\,(1972)\nolinebreak [2]\,438 and 675.

\bibitem{DGLAPb}
L.~Lipatov, {\em Sov. J. Nucl. Phys.} {\bf 20}\nolinebreak
  [2]\,(1975)\nolinebreak [2]\,94.

\bibitem{DGLAPc}
G.~Altarelli, G.~Parisi, {\em Nucl. Phys. {\bf B}} {\bf 126}\nolinebreak
  [2]\,(1977)\nolinebreak [2]\,298.

\bibitem{DGLAPd}
Y.~Dokshitser, {\em Sov. Phys. JETP} {\bf 46}\nolinebreak
  [2]\,(1977)\nolinebreak [2]\,641.

\bibitem{Ingelman_LEPTO65}
G.~Ingelman, A.~Edin, J.~Rathsman, {\em Comp. Phys. Comm.} {\bf
  101}\nolinebreak [2]\,(1997)\nolinebreak [2]\,108.

\bibitem{Jetsetc}
T.~Sj\"ostrand, {\em Comp. Phys. Comm.} {\bf 82}\nolinebreak
  [2]\,(1994)\nolinebreak [2]\,74.

\bibitem{Herwig}
G.~Marchesini et~al., {\em Comp. Phys. Comm.} {\bf 76}\nolinebreak
  [2]\,(1992)\nolinebreak [2]\,465, hep-ph/9912396.

\bibitem{Herwig54}
B.~Webber, Herwig 5.4,  in {\em Proc.\ of the Workshop on Physics at HERA Vol.
  3, 1354}, edited by W.~Buchm\"uller, G.~Ingelman (1991).

\bibitem{RAPGAP206}
H.~Jung, {\em \mbox{T}he RAPGAP Monte Carlo for Deep Inelastic Scattering,
  version 2.08}, Lund University, 1999,
  \verb+http://www-h1.desy.de/~jung/rapgap.html+.

\bibitem{CCFMa}
M.~Ciafaloni, {\em Nucl. Phys. {\bf B}} {\bf 296}\nolinebreak
  [2]\,(1988)\nolinebreak [2]\,49.

\bibitem{CCFMb}
S.~Catani, F.~Fiorani, G.~Marchesini, {\em Phys. Lett. {\bf B}} {\bf
  234}\nolinebreak [2]\,(1990)\nolinebreak [2]\,339.

\bibitem{CCFMc}
S.~Catani, F.~Fiorani, G.~Marchesini, {\em Nucl. Phys. {\bf B}} {\bf
  336}\nolinebreak [2]\,(1990)\nolinebreak [2]\,18.

\bibitem{CCFMd}
G.~Marchesini, {\em Nucl. Phys. {\bf B}} {\bf 445}\nolinebreak
  [2]\,(1995)\nolinebreak [2]\,49.

\bibitem{PYTHIAPSa}
T.~Sj\"ostrand, {\em Phys. Lett. {\bf B}} {\bf 157}\nolinebreak
  [2]\,(1985)\nolinebreak [2]\,321.

\bibitem{PYTHIAPSb}
M.~Bengtsson, T.~Sj\"ostrand, M.~van Zijl, {\em Z. Phys. {\bf C}} {\bf
  32}\nolinebreak [2]\,(1986)\nolinebreak [2]\,67.

\bibitem{MarchWebbBE}
G.~Marchesini, B.~Webber, {\em Nucl. Phys. {\bf B}} {\bf 310}\nolinebreak
  [2]\,(1988)\nolinebreak [2]\,461.

\bibitem{SMALLXa}
G.~Marchesini, B.~Webber, {\em Nucl. Phys. {\bf B}} {\bf 349}\nolinebreak
  [2]\,(1991)\nolinebreak [2]\,617.

\bibitem{SMALLXb}
G.~Marchesini, B.~Webber, {\em Nucl. Phys. {\bf B}} {\bf 386}\nolinebreak
  [2]\,(1992)\nolinebreak [2]\,215.

\bibitem{BFKLa}
E.~Kuraev, L.~Lipatov, V.~Fadin, {\em Sov. Phys. JETP} {\bf 44}\nolinebreak
  [2]\,(1976)\nolinebreak [2]\,443.

\bibitem{BFKLb}
E.~Kuraev, L.~Lipatov, V.~Fadin, {\em Sov. Phys. JETP} {\bf 45}\nolinebreak
  [2]\,(1977)\nolinebreak [2]\,199.

\bibitem{BFKLc}
Y.~Balitskii, L.~Lipatov, {\em Sov. J. Nucl. Phys.} {\bf 28}\nolinebreak
  [2]\,(1978)\nolinebreak [2]\,822.

\bibitem{LDCa}
B.~Andersson, G.~Gustafson, J.~Samuelsson, {\em Nucl. Phys. {\bf B}} {\bf
  467}\nolinebreak [2]\,(1996)\nolinebreak [2]\,443.

\bibitem{LDCb}
B.~Andersson, G.~Gustafson, H.~Kharraziha, J.~Samuelsson, {\em Z. Phys. {\bf
  C}} {\bf 71}\nolinebreak [2]\,(1996)\nolinebreak [2]\,613.

\bibitem{LDCc}
G.~Gustafson, H.~Kharraziha, L.~L\"onnblad, The LCD Event Generator,  in {\em
  Proc. of the Workshop on Future Physics at HERA}, edited by A.~\mbox{De
  Roeck}, G.~Ingelman, R.~Klanner (1996), p.\ 620.

\bibitem{LDCd}
H.~Kharraziha, L.~L\"onnblad, {\em JHEP} {\bf 03}\nolinebreak
  [2]\,(1998)\nolinebreak [2]\,006.

\bibitem{ForshawSabioVera98}
J.~R. Forshaw, A.~{Sabio Vera}, {\em Phys. Lett.} {\bf B440}\nolinebreak
  [2]\,(1998)\nolinebreak [2]\,141.

\bibitem{Webber98}
B.~R. Webber, {\em Phys. Lett.} {\bf B444}\nolinebreak [2]\,(1998)\nolinebreak
  [2]\,81.

\bibitem{Salam99}
G.~P. Salam, {\em JHEP} {\bf 03}\nolinebreak [2]\,(1999)\nolinebreak [2]\,009.

\bibitem{Schmidt}
C.~R. Schmidt, {\em Phys. Rev. Lett.} {\bf 78}\nolinebreak
  [2]\,(1997)\nolinebreak [2]\,4531.

\bibitem{OrrStirlingA}
L.~H. Orr, W.~J. Stirling, {\em Phys. Rev.} {\bf D56}\nolinebreak
  [2]\,(1997)\nolinebreak [2]\,5875.

\bibitem{OrrStirlingB}
L.~H. Orr, W.~J. Stirling, {\em Phys. Lett.} {\bf B429}\nolinebreak
  [2]\,(1998)\nolinebreak [2]\,135.

\bibitem{Salam}
G.~Bottazzi, G.~Marchesini, G.~Salam, M.~Scorletti, {\em JHEP} {\bf
  12}\nolinebreak [2]\,(1998)\nolinebreak [2]\,011, hep-ph/9810546.

\bibitem{NLLFL}
V.~S. Fadin, L.~N. Lipatov, {\em Phys. Lett.} {\bf B429}\nolinebreak
  [2]\,(1998)\nolinebreak [2]\,127.

\bibitem{NLLCC}
M.~Ciafaloni, G.~Camici, {\em Phys. Lett.} {\bf B430}\nolinebreak
  [2]\,(1998)\nolinebreak [2]\,349.

\bibitem{BMSSunpublished}
G.~Bottazzi, G.~Marchesini, G.~Salam, M.~Scorletti, {\em unpublished.}

\bibitem{Martin_Sutton}
J.~Kwiecinski, A.~Martin, P.~Sutton, {\em Phys. Rev. {\bf D}} {\bf
  52}\nolinebreak [2]\,(1995)\nolinebreak [2]\,1445.

\bibitem{LDCgoodLL}
B.~Andersson, G.~Gustafson, H.~Kharraziha, L.~L\"onnblad, {\em private
  communication.}

\bibitem{smallx_f2}
H.~Jung, CCFM prediction for $F_2$ and forward jets at HERA,  in {\em
  Proceedings of Workshop on Deep Inelastic Scattering and QCD (DIS 99)}
  (\mbox{DESY}, Zeuthen, 1999), hep-ph/9905554.

\bibitem{CASCADE}
H.~Jung, CCFM prediction on forward jets and $F_2$: parton level predictions
  and a new hadron level Monte Carlo generator {\sc Cascade},  in {\em
  Proceedings of the Workshop on Monte Carlo generators for HERA physics},
  edited by A.~Doyle, G.~Grindhammer, G.~Ingelman, H.~Jung (\mbox{DESY},
  Hamburg, 1999), hep-ph/9908497.

\bibitem{off_shell_me}
S.~Catani, M.~Ciafaloni, F.~Hautmann, {\em Nucl. Phys. {\bf B}} {\bf
  366}\nolinebreak [2]\,(1991)\nolinebreak [2]\,135.

\bibitem{Jetsetnew}
T.~Sj\"ostrand, {\em Comp. Phys. Comm.} {\bf 82}\nolinebreak
  [2]\,(1994)\nolinebreak [2]\,74.

\bibitem{H1_F2_1996}
\mbox{H1 Collaboration, S. Aid et al.}, {\em Nucl. Phys. {\bf B}} {\bf
  470}\nolinebreak [2]\,(1996)\nolinebreak [2]\,3.

\bibitem{H1_fjets_data}
\mbox{H1 Collaboration, C. Adloff et al.}, {\em Nucl. Phys. {\bf B}} {\bf
  538}\nolinebreak [2]\,(1999)\nolinebreak [2]\,3.

\bibitem{ZEUS_fjets_data}
\mbox{ZEUS Collaboration}; J. Breitweg~et al., {\em Eur. Phys. J. {\bf C}} {\bf
  6}\nolinebreak [2]\,(1999)\nolinebreak [2]\,239.

\bibitem{ZEUS_fjets_pt2/q2}
\mbox{ZEUS Collaboration}; J. Breitweg~et al., {\em Phys. Lett. {\bf B}} {\bf
  474}\nolinebreak [2]\,(1999)\nolinebreak [2]\,223.

\bibitem{Kuhlena}
M.~Kuhlen, {\em Phys. Lett. {\bf B}} {\bf 382}\nolinebreak
  [2]\,(1996)\nolinebreak [2]\,441, hep-ph/9606246.

\bibitem{Kuhlenb}
M.~Kuhlen, High $p_t$ particles in the forward region at HERA,  in {\em Proc.
  of the Workshop on Future Physics at HERA}, edited by A.~\mbox{De Roeck},
  G.~Ingelman, R.~Klanner (\mbox{DESY}, Hamburg, 1996), p.\ 606,
  hep-ex/9610004.

\bibitem{H1_ptspectra_data}
\mbox{H1 Collaboration, C. Adloff et al.}, {\em Nucl. Phys. {\bf B}} {\bf
  485}\nolinebreak [2]\,(1997)\nolinebreak [2]\,3.

\bibitem{ZEUS_dstar}
\mbox{ZEUS Collaboration}; J. Breitweg~et al., {\em Eur. Phys. J. {\bf C}} {\bf
  6}\nolinebreak [2]\,(1999)\nolinebreak [2]\,67.

\bibitem{H1_bbar}
\mbox{H1} Collaboration; C. Adloff~et al., {\em Phys. Lett. {\bf B}} {\bf
  467}\nolinebreak [2]\,(1999)\nolinebreak [2]\,156.

\bibitem{ZEUS_bbar}
\mbox{ZEUS} Collaboration; J. Breitweg~et al., {\em \mbox{submitted to} Eur.
  Phys. J. {\bf C}}, \mbox{DESY 00-166} hep-ex/0011081.

\end{thebibliography}
\end{document}